\documentclass[twocolumn]{aastex62}
\usepackage{graphicx}
\usepackage{graphics}
\usepackage{natbib} 

\newcommand\aastex{AAS\TeX}

\newcommand{\diamonds}{\textsc{D\large{iamonds}}}

\received{April 18, 2018}
\accepted{May 22, 2018}
\submitjournal{ApJ}

\shorttitle{\aastex\ The rotational shear layer inside the red giant star KIC~4448777}
\shortauthors{Di Mauro et al.}

\begin{document}

\title{The rotational shear layer inside the early red-giant star KIC~4448777}

\correspondingauthor{Maria Pia Di Mauro}
\email{maria.dimauro@inaf.it}

\author{Maria Pia Di Mauro }
\affil{INAF-IAPS, Istituto di Astrofisica e Planetologia Spaziali\\
Via del Fosso del Cavaliere 100 \\
00133 Roma, Italy}

\author{Rita Ventura}
\affil{INAF-Astrophysical Observatory of Catania\\
Via S. Sofia 78\\
95123 Catania, Italy}

\author{Enrico Corsaro}
\affil{INAF-Astrophysical Observatory of Catania\\
Via S. Sofia 78\\
95123 Catania, Italy}

\author{Bruno Lustosa De Moura}
\affil{Universidade Federal do R. G. do Norte - Natal, UFRN, Brazil}
\affil{Instituto Federal do R. G. do Norte - Natal, IFRN, Brazil}



\begin{abstract}

We present the asteroseismic study of the early red-giant star KIC~4448777, complementing and integrating a previous work  \citep[][Paper I]{dimauro2016}, aimed at characterizing the dynamics of its interior by analyzing the overall set of data collected by the {\it Kepler} satellite during the four years of its first nominal mission.  We adopted the Bayesian inference code \diamonds\,\,\citep{Corsaro14} for the peak bagging analysis and asteroseismic splitting inversion methods to derive the internal rotational profile of the star. The detection of new splittings of mixed modes, more concentrated in the very inner part of the helium core, allowed us to reconstruct the angular velocity profile deeper into the interior of the star and to disentangle the details better than in Paper I: the helium core rotates almost rigidly about 6 times faster than the convective envelope, while part of the hydrogen shell seems to rotate at a constant velocity about 1.15 times lower than the He core. In particular, we studied the internal shear layer between the fast-rotating radiative interior and the slow convective zone and we found that it lies partially inside the hydrogen shell above $r \simeq 0.05R$ and extends across the core-envelope boundary. Finally, we theoretically explored the possibility for the future to sound the convective envelope in the red-giant stars and we concluded that the inversion of a set of splittings with only low-harmonic degree $l\leq 3$, even supposing a very large number of modes, will not allow to resolve the rotational profile of this region in detail.

\end{abstract}

\keywords{stars: oscillations, stars: AGB, stars: interiors, stars: individual (KIC~4448777), stars: solar-type, stars: rotation}
~\\

\section{Introduction} \label{sec:intro}
Although stellar rotation has long been recognized as an important mechanism capable of strongly  affecting stellar structure and evolution  \citep[e.g.,][]{maeder2009},  a clear understanding of the processes that transport and redistribute angular momentum in stellar interiors at all phases of the evolution is still lacking. As a consequence the radial differential rotation profiles derived from evolutionary models are in general poorly characterized. 

Several classes of mechanisms transporting angular momentum in stars have been proposed  and their effects extensively evaluated: purely hydrodynamical instabilities, such as shear turbulence and meridional circulation \citep[see the review by][]{maeder2012} also recently implemented in the hypothesis of shellular rotation \citep{zahn1992} and enhanced viscosity \citep{eggenberger2017}, internal gravity- waves excited at the edge of the convective regions \citep{charbonnel2005} and magnetic mechanisms, such as magnetic torques \citep{gough1998, spruit1999, spruit2002, spada2010}.  In all the considered cases, due to a poor core-envelope coupling,  the resulting internal rotation profiles are characterized by very fast-spinning cores and slowly rotating envelopes with a shear layer in between \citep[e.g.,][]{eggenberger2012}.

In recent years, asteroseismology with the {\it Kepler} satellite \citep{borucki2010} has opened a new era, providing us with unprecedented tools for studying the internal rotation profile and its temporal evolution in a large sample of stars with different mass and evolutionary stages.  
The detection of pulsation modes with mixed g-p character in subgiant and red-giant stars,
has strongly helped to progress in modeling the angular momentum transport mechanisms operating in stellar interiors.
In particular, rotational splittings of oscillation frequencies have shown that the cores of red giants are rotating  from 5  to 20 times faster than  the envelopes \citep{beck2012, deheuvels2012, mosser2012a, deheuvels2014, dimauro2016, Triana17}. This unexpected found, at odds with the current theoretical models that predict rotation rates at least 10 times higher than observed  \citep{eggenberger2012, marques2013,  ceillier2013, cantiello}, implies that a still unknown process (or processes) redistributing  angular momentum inside the stars must be operating at certain phases of the evolution and calls for new efforts both to implement more efficient physical mechanisms into stellar evolution codes and to get stringent constraints from observations. 
 
The notion of a shear layer at the bottom of the convection zone
had been present in evolutionary models for some time prior to its observational discovery.
The existence of the 'tachocline' \citep{spiegel}, the layer of strong radial shear around the base of the convective zone, 
 was firstly proved in the Sun by \cite{brown}. 
Its discovery offered a
solution to the puzzle of the apparent absence of a radial gradient of rotation
in the convection zone that could drive a solar dynamo, leading to speculation
that the dynamo must operate in the 'tachocline' region instead of in the bulk of
the convection zone.

The characterization of the shear layer between fast-rotating core and slow envelope also in stars more evolved than the Sun is a challenging objective for asteroseismology \citep{beck2012, deheuvels2014, dimauro2016}. Answering this question would strongly constrain the dominant
mechanisms of angular momentum transport in red giants and provides an extremely powerful tool to discriminate among different rotational models, as it has been very recently demonstrated by \citet{klion2017}. They considered two classes of possible theoretical profiles, one where the differential rotation is concentrated just outside the hydrogen burning shell \citep[e.g.,][]{eggenberger2012} and another one where the differential rotation resides in the convective envelope \citep[e.g.,][]{kissin2015}. 
Here we try to contribute to distinguish between these two different models, complementing previous studies on the red giant KIC~4448777 \citep[][hereafter Paper I]{dimauro2016}.  

The star KIC~4448777, located at the beginning of the ascending red-giant branch, has been observed by the {\it Kepler} satellite for more than four years, during its first nominal mission. More than two years of observations of the star have been analyzed in Paper I. At that time the authors identified 14 rotational splittings of mixed modes and characterized its internal rotational profile using inversion techniques previously applied with success to helioseismic data \citep[e.g.,][]{thompson96, schou1998, paterno1996, dimauro1998}.  They were able to establish that the core of the star rotates rigidly from 8 to 17 times faster than the surface and  provided evidence for a discontinuity in the inner stellar rotation located between  the helium core and part of the hydrogen-burning shell. However, due to the modest number of dipolar mode splittings detected in the spectrum of the star it was not possible to infer the complete internal rotational profile  and to distinguish between a smooth or sharp gradient in the angular velocity. In this paper we analyze all the available data collected by the {\it Kepler} satellite on KIC~4448777, i.e. more than four years of uninterrupted  observations, providing a formal frequency resolution of about 8~nHz, greatly improved with respect to the value of 15~nHz relative to the data-set analyzed in Paper I.
The paper is organized as follows:  Section 2 describes the method adopted to analyze the oscillation spectrum and to identify the modes, the frequencies and the related splittings. Section 3 presents the details of the asteroseismic inversion carried out to infer the rotational profile of the star, describes the evolutionary models constructed to best fit the atmospheric and asteroseismic constraints and discusses the surface term correction.  Section 4 presents the results of the asteroseismic inversion.  In Section 5  the  possibility of studying the angular velocity profile inside the convective envelope is discussed. Section 6 summarizes the results and draws the conclusions.

\section{Data analysis} \label{sec:data}

For the asteroseismic analysis we have used the near-continuous photometric time
series obtained by {\it Kepler} in long-cadence mode (time sampling of
29.4 min).  This light curve spans more than four years corresponding to
observing quarters Q0-17, providing a formal frequency resolution of 8~nHz.
We used the so-called PDC-SAP (pre data conditioning - simple aperture photometry) light curve \citep{jenkins2010} corrected for
instrumental trends and discontinuities as described by \citet{garcia11}.

\begin{figure*}
\centering
\includegraphics[width=16cm]{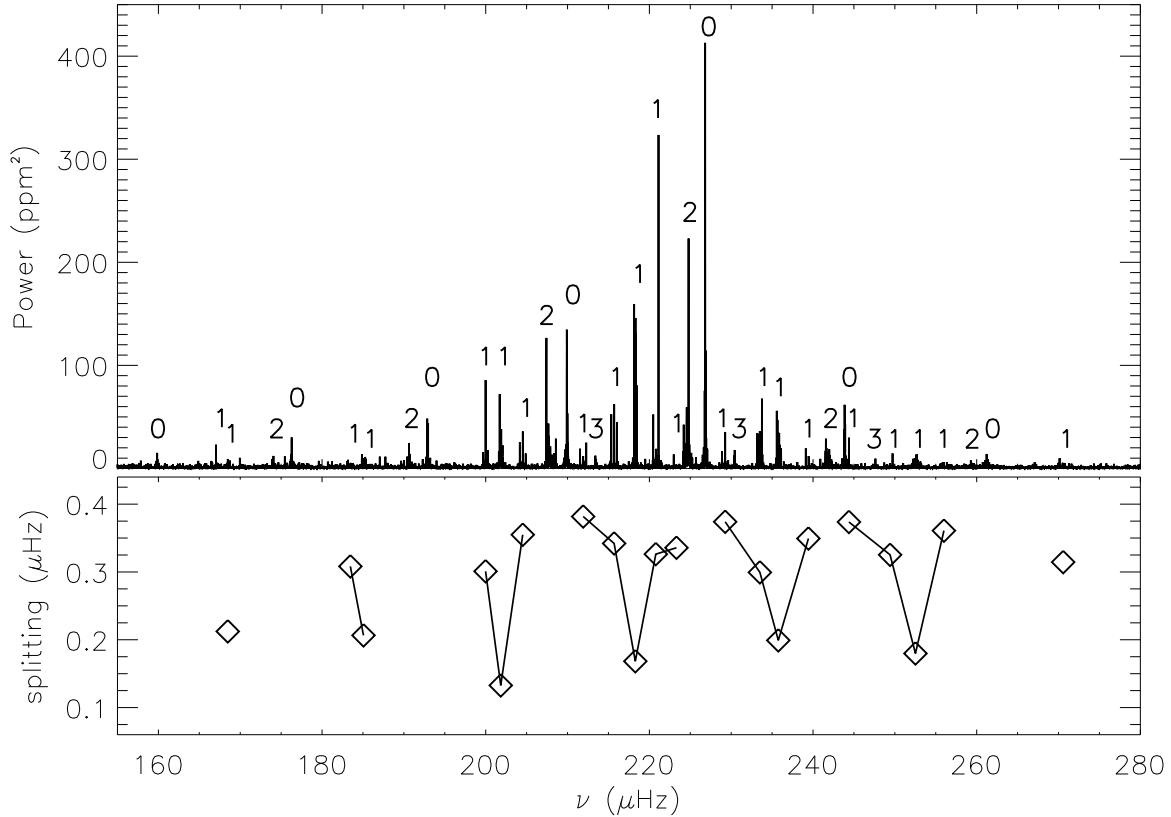}
\caption{Panel a: Observed frequency spectrum of KIC~4448777. The harmonic
degrees of the modes ($l = 0, 1, 2, 3$) are indicated. Multiplets due to rotation are
visible for $l=1$. Panel b: Observed rotational splittings for $l=1$ modes.}
\label{fig:spectrum}
\end{figure*}

The power spectrum of the light curve, obtained by adopting the IDL Lomb-Scargle algorithm \citep{lomb1976,scargle1982},  shows a 
clear power excess in the range $(170-260)\, \mu$Hz
(Fig. \ref{fig:spectrum}) due to radial modes, with the comb-like pattern typical of the solar-like p-mode
oscillations, and non-radial modes, particularly those of spherical degree $l=1$,
modulated by the mixing with g modes. 
For analyzing the power spectrum we used the Bayesian inference code \diamonds\,\,\citep{Corsaro14}. In particular, to fit and extract the properties of the individual oscillation modes we followed the procedure presented by \cite{Corsaro15} for the peak bagging analysis of a red giant star, which we summarize below. First, we estimated the level of the background signal, comprising two granulation-related components, one component originating from signal at low-frequency (e.g. rotational modulation, activity, and super-granulation), a Gaussian envelope containing the region of the oscillations, and a white noise component (see Eq. 1 in \citealt{Corsaro15}). Second, we fixed the resulting background level and replaced the Gaussian envelope with a detailed peak bagging model including a mixture of resolved and unresolved peak profiles (see Eqs.7-8 in \citealt{Corsaro15}), according to the lifetime of each mode. We used Lorentzian profiles for the fit of all the radial and quadrupole modes, while part of the dipole modes were fitted using a $\mbox{sinc}^2$ profile, as explained by \cite{Corsaro15}. The oscillation modes were identified using the Tassoul's asymptotic relation for p modes \citep{Tassoul80} and the asymptotic relation for dipole mixed modes \citep{Mosser12b}. We also computed detection probabilities using the Bayesian model comparison to test the significance of those peaks with low signal-to-noise ratio in the power spectrum.

\begin{deluxetable*}{cccccccc}
\tabletypesize{\scriptsize}
\tablecolumns{8}
\tablewidth{0pt}
\tablecaption{Observed frequencies  and rotational splittings for KIC~4448777 as derived using \diamonds. Median values and 68\% Bayesian credible intervals are considered.}
\tablehead{\colhead{$l$} & \colhead{$m$} & \colhead{$\nu_{n,l} (\mu{\mathrm Hz})$}  & \colhead{$\delta\nu_{n,l} (\mu {\mathrm Hz})$} & 
\colhead{$l$} & \colhead{$m$} & \colhead{$\nu_{n,l} (\mu{\mathrm Hz})$}  & \colhead{$\delta\nu_{n,l} (\mu {\mathrm Hz})$}}
\startdata
 0  & 0  &  $159.8416 \pm 0.0276$ &   & 1 & -1~ & $222.9770 \pm 0.0011$ &  \\[1pt]
 0  & 0  & $  176.2709 \pm 0.0132$ &   & 1 & 0 &  $   223.3127 \pm 0.0027$ &  { \boldmath $0.3357 \pm 0.0029$}\\[1pt]
 0  & 0  & $ 192.9170 \pm 0.0132$ &  & 1  & -1~ &  $ 228.8866 \pm 0.0007$ & \\[1pt] 
 0 & 0 & $    209.9247 \pm 0.0119$ &   & 1 & 0 &  $ 229.2647 \pm 0.0011$ & $ 0.3740 \pm 0.0008$ \\[1pt]
 0 & 0 & $    226.8168 \pm 0.0059$ &  & 1 & 1 &  $  229.6346 \pm 0.0014$ &  \\[1pt] 
 0 & 0 & $    243.8681 \pm 0.0116$ &  & 1  & -1~ & $    233.2020 \pm 0.0128$ &  \\[1pt]  
 0  & 0 & $    261.1653 \pm0.0397$ &  & 1  & 0 & $    233.5070 \pm 0.0064$ &  $0.2994 \pm 0.0077$ \\[1pt]
1 & 0 &  $    167.0459 \pm 0.0011$ &   & 1 & 1 & $    233.8009 \pm 0.0085$ &  \\[1pt]   
 1 & -1~ & $  168.2686 \pm 0.0021$\tablenotemark{a} &   &  1 & -1~ & $ 235.5743 \pm 0.0059$ & \\[1pt]   
1 & 0 &     $168.4809 \pm 0.0084$\tablenotemark{a} &  {\boldmath $0.2123 \pm 0.0087$} &  1  & 0 & $ 235.7545 \pm 0.0071$ &  $0.1991 \pm 0.0072$ \\[1pt]
1  & 0  & $   170.0257 \pm 0.0200$\tablenotemark{a} &  & 1  & 1  & $ 235.9725 \pm 0.0131$ &   \\[1pt]
  1  & -1~ &  $   183.1591 \pm 0.0027$ & &  1 &  -1~ & $    239.1292 \pm 0.0178$ & \\[1pt]  
1 & 0 &  $    183.4617 \pm 0.0034$ &   { \boldmath $0.3082 \pm  0.0016$} &  1  & 0  & $    239.4356 \pm 0.0092$ & $0.3493 \pm 0.0095$   \\[1pt] 
 1 & 1 &  $    183.7756 \pm 0.0018$ &  &  1 &  1 & $ 239.8279 \pm 0.0071$ & \\[1pt]   
 1 & -1~ &  $    184.8626 \pm 0.0045$ &   &  1 & 0 &  $  244.3870 \pm 0.0007$ &  { \boldmath $ 0.3737 \pm 0.0013$} \\[1pt] 
1  & 0  & $  185.0691 \pm 0.0122$ & $0.2065  \pm 0.0080$ & 1 & 1 &  $    244.7607 \pm 0.0011$ &  \\[1pt] 
 1  &  1 & $   185.2794 \pm 0.0154$ & &   1  & -1~  &  $    249.0567\pm 0.0420$\tablenotemark{a} & \\[1pt]  
 1 & -1~ & $  199.6719 \pm 0.0121$ &     &  1 & 0 &  $249.4165\pm 0.0024$\tablenotemark{a}  & $0.3253 \pm 0.0211$ \\[1pt]
1 & 0 &  $    199.9857 \pm 0.0007$ & $0.3010 \pm 0.0077$ & 1  & 1 &  $    249.7074 \pm 0.0029$ &  \\[1pt]
1  & 1  & $  200.2739 \pm 0.0096$ & &  1  &  -1~ & $    252.3030 \pm 0.0300$ & \\[1pt]  
1 & -1~  & $    201.7064 \pm 0.0047$ & &     1 &  0 & $    252.5167 \pm 0.0357$ &  {\boldmath $0.1798 \pm 0.0342$} \\[1pt]
1& 0& $    201.8391 \pm 0.0129$ &  $0.1327 \pm 0.0036$ &  1 & 1 & $    252.6626 \pm  0.0615$ & \\[1pt]
1 &1 &  $  202.1270 \pm 0.0054$\tablenotemark{a} &    &   1 & -1~  & $    255.6352 \pm 0.0350$\tablenotemark{a} &  \\[1pt] 
 1 & -1~ &  $    204.1804 \pm 0.0007$ &   &  1 & 0  &   $255.9972  \pm 0.0207$\tablenotemark{a} & { \boldmath $0.3609 \pm 0.0176$} \\[1pt]   
1 & 0 &  $    204.5278 \pm 0.0012$  & $0.3549 \pm 0.0008$  &  1  & 1 & $256.3571 \pm 0.0043$\tablenotemark{a} & \\[1pt]   
1 & 1 &  $    204.8903 \pm 0.0014$ & &  1 & 1 &   $    267.9713 \pm 0.0023$\tablenotemark{a} &  \\[1pt]    
1 & 1 &  $    208.5712 \pm 0.0118$\tablenotemark{a} &   &   1  &  -1~ & $    270.1864 \pm 0.0368$\tablenotemark{a} &   \\[1pt] 
1 & -1~ & $    211.5236 \pm 0.0021$ &  &   1  &  0 &   $270.5683 \pm 0.0259$\tablenotemark{a} & { \boldmath $0.3145 \pm 0.0303$}  \\[1pt]  
1 & 0 &  $    211.9101 \pm 0.0012$ & $0.3818 \pm 0.0016$ &   1 & 1 &   $    270.8155 \pm 0.0482$\tablenotemark{a} & \\[1pt]
1 & 1 &  $    212.2872 \pm 0.0023$ &  &  2 & 0 & $173.9291 \pm 0.0272$ &\\[1pt]   
 1 & -1~  & $    215.3498 \pm 0.0007$ &  &  2 & 0 & $190.6347 \pm 0.0176$    &\\[1pt] 
 1 & 0 &  $    215.6962 \pm 0.0007$ &  $0.3421 \pm 0.0005$ &   2 & 0 &  $207.6233 \pm 0.0266$  &\\[1pt]
1 & 1 &  $    216.0341 \pm 0.0007$ &   &   2 & 0 & $  224.7066 \pm 0.0102$  &\\[1pt]
1 & -1~ &  $    218.1359 \pm 0.0006$  & & 2 & 0 & $  241.7668 \pm 0.0236$ &\\[1pt]
1 &  0 & $    218.2847 \pm 0.0132$ & $0.1683 \pm 0.0045$ &   2 & 0 &  $  259.2743 \pm 0.0826$  &\\[1pt]
1  &  1 & $    218.4726 \pm 0.0089$  & &  3 & 0 & $ 213.4207 \pm 0.0150$ &\\[1pt] 
1 & -1~ &  $    220.4658 \pm 0.0011$  & &  3 & 0 & $  230.4102 \pm 0.0080$  &\\[1pt] 
 1 & 0 &  $    220.7879 \pm 0.0179$ & $0.3263 \pm 0.0006$ &   3 & 0 & $  247.6574 \pm 0.0351$  &\\[1pt] 
1 & 1 &  $    221.1185 \pm 0.0007$ &  &   & &   &\\[1pt] 
\enddata
 \tablenotetext{a}{Detection probability $\leq 0.99$, (see \citealt{Corsaro14,Corsaro15} for more details).\\ New splittings, detected in the complete four-year data set and unrevealed in the spectrum analyzed in Paper I,  are reported in boldface.}
\label{tab:frequencies}
\end{deluxetable*}



Table~\ref{tab:frequencies} lists the final set of 77 individual frequencies, including the multiplets due to rotation for the $l=1$ modes, together with their
uncertainties, corresponding to the values obtained using \diamonds\,\,for radial and quadrupole modes, their spherical degree and azimuthal order, and the rotational splittings for 20 dipole modes. The peaks with a detection probability below the limiting threshold proposed by \cite{Corsaro15} are marked.

In addition, following the method used by \cite{Corsaro17}, we estimated the large separation $\Delta\nu$ using a linear Bayesian fit over the asymptotic relation for the p modes of the central radial mode frequencies, closes to $\nu_\mathrm{max}$. Our result, with its 68\% Bayesian credible interval, is given in Table \ref{tab:fitted}.

\section{Asteroseismic inversion}

The internal angular velocity of KIC~4448777 has been probed
by asteroseismic inversion of the 20 rotational splittings obtained from the complete four-year observational data set, by solving the following integral equations:
\begin{equation}
\delta\nu_{i} = \int_{0}^{R} {\cal K}_{i}(r) \frac{\Omega(r)}{2 \pi}\, dr +\epsilon_{i}\;.
\label{eq:rot}
\end{equation}
Equation \ref{eq:rot}, which relates the set of observed rotational splittings $\delta\nu_{i}$, with uncertainties $\epsilon_{i}$ for the modes $i=(n,l)$, to  the internal rotational profile $\Omega(r)$, is
derived from the application of a standard perturbation theory to an
equilibrium stellar structure model, in the hypothesis of slow rotation \citep{gough1981} and when the rotation is assumed to be independent of latitude.

The functions
${\cal K}_{n,l}(r)$ are  the mode kernels  calculated on the unperturbed eigenfunctions for modes ($n,l$) and other physical quantities of the stellar model which best reproduces all the observational constraints of the star  (see Sects. 4 and 5 of Paper I for details).

\subsection{Evolutionary models and surface term correction}

In order to select the best structure model of KIC~4448777,
we reconsidered all the theoretical models produced by using the ASTEC evolution code \citep{chris2008a} for the purpose of Paper I. This was necessary because the new set of observed frequencies is characterized by additional modes and different observational errors with respect to the one used in Paper I. 
  The goodness of the fits between the observed frequencies and the theoretical ones computed for the models has been
  evaluated by calculating for the set of $N$ modes
  the total $\chi^2$
   between the observed $\nu^{obs}_i$ and the corrected model frequencies  $\nu^{mod}_i$  as:
  \begin{equation}
  \chi^2=\frac{1}{N}\sum_{i=1}^N\left(\frac{\nu^{obs}_i-\nu^{mod}_{i}}{\epsilon_i}\right)^2.
  \end{equation}
    
Unfortunately our difficulty to properly model the near surface layers of the stars with outer convective envelopes makes the selection of the best model not so straightforward, since theoretical models of stars are inevitably affected by errors. In fact, there are several physical
mechanisms  such as convective flux, non-adiabatic properties, interaction with oscillations, equation of state that are still poorly known. These physical processes have been - so far - inadequately  investigated, and in most cases even neglected in the stellar model computations.
Thus, the theoretical frequencies are generally calculated in the adiabatic approximation, which is certainly inappropriate  in the near surface region, where the thermal time scale becomes comparable with the oscillation period. In order to overcome the lack of a proper theory for the description of oscillations in the upper surface layers,
it is common procedure to correct the theoretical frequencies by a term found empirically by analyzing the differences between observed and theoretical frequencies and firstly introduced by \citet{kjeldsen2008}.

 Here we applied 
the surface-effect correction following the approach proposed by \citet{ballgizon2014}, which has been proved by \citet{ballgizon2017} and \citet{schmittbasu}  to work much better for evolved stars than the  approach proposed by \citet{kjeldsen2008}, which we used in Paper I. Another  valid empirical surface correction was also proposed by 
\citet{Sonoi}, whose prescription was not adopted here.

The surface correction proposed by \citet{ballgizon2014} includes
 two different terms
 and it is defined by  the following equation:
\begin{equation}
\nu^{mod}_{n,l}=\nu_{n,l}+a_3\frac{1}{E_{n,l}}\left(\frac{\nu_{n,l}}{\nu_{ac}}\right)^3+
a_{-1}\frac{1}{E_{n,l}}\left(\frac{\nu_{n,l}}{\nu_{ac}}\right)^{-1}\, ,
\label{corr}
\end{equation}
where  $\nu^{mod}_{n,l}$ are the corrected frequencies, $E_{n,l}$ is the inertia of the given mode normalized by the inertia at the surface, $a_{-1}$ and $a_3$ are parameters found by trying to minimize the differences between observed and model frequencies, 
 $\nu_{ac}$ is the acoustic cutoff frequency calculated such as:
\begin{equation}
\nu_{ac}=\nu_{ac,\odot}\frac{g}{g_{\odot}}\left(\frac{T_{\mathrm eff}}{T_{\mathrm eff, \odot}}\right)^{-1/2}
\end{equation}
where $\nu_{ac,\odot}=5000\, \mu Hz$ and  $T_{{\mathrm eff}, \odot}=5777\,K$ are the values measured  for the Sun. 
 
In Paper I, it was not possible to distinguish which of two selected models, characterized by different parameters, was the closest to KIC~4448777.  Table  \ref{tab:fitted} reports for those two models, the mass, the effective temperature, the gravity, the surface radius, the luminosity, the initial metallicity $Z_i$, the hydrogen abundance $X_i$, the iron abundance $[\mathrm{Fe/H}]$, the location of the base of the convective region $r_{cz}$, the  extent of the He core $r_{He}$  and the large separation $\Delta \nu$ calculated by linear fit over the asymptotic relation for the corrected radial mode frequencies.
The initial heavy-element mass fraction $Z_i$
has been calculated from the spectroscopically observed iron abundance using the relation [Fe/H]$=\log(Z/X)-\log(Z/X)_{\odot}$, where $(Z/X)$ is the value at
the stellar surface and the solar value is taken from \citep{GN93}.
All the theoretical values are compared with the observed ones in Table \ref{tab:fitted}.

\begin{figure*}
\begin{center}
\includegraphics[width=14cm]{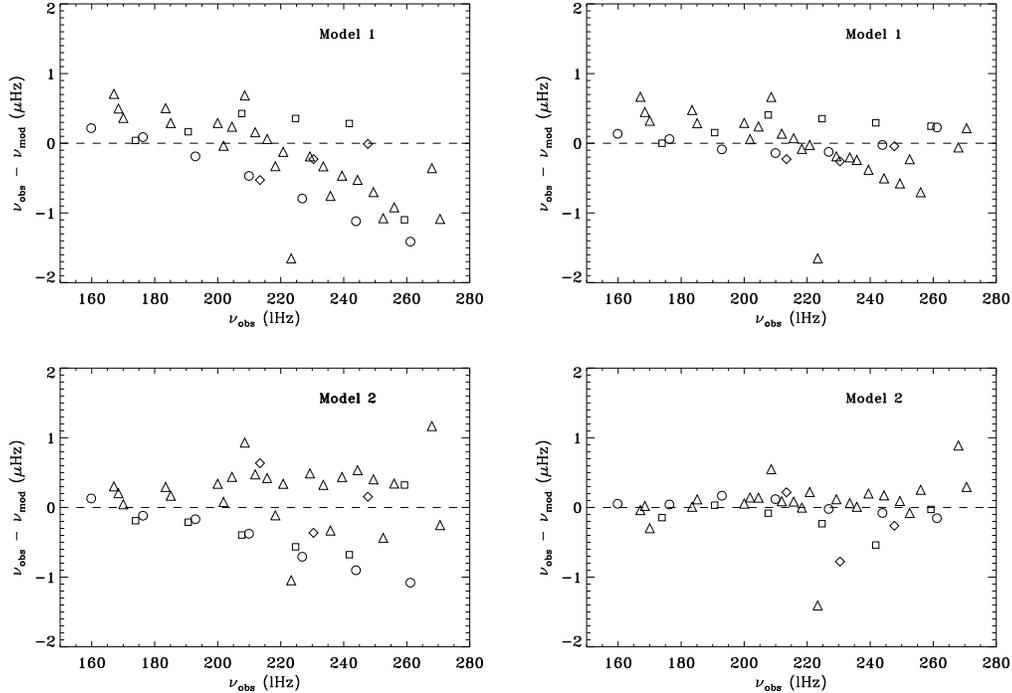}
\caption{Differences between modelled and observed frequencies plotted against observed frequencies for Model~1 (upper row) and Model~2 (lower row) without correction (panels on the left) and with surface corrections (panels on the right). Circles are used for modes with $l=0$, triangles for $l=1$, squares for $l=2$, diamonds for $l=3$.}
\end{center}
\label{fig:surfcorr}
\end{figure*}

Figure \ref{fig:surfcorr} shows the difference between modelled and observed
frequencies for Model~1 and Model~2 with and without surface corrections.
As it is clearly evident,  Model~2, with surface-correction effects included, provides the best fit of the observed frequencies of the star.  The use of a larger set of data with respect to Paper I greatly improved the fit,  allowing us to definitely discriminate between the two models.  
 It is not surprising the fact that the Model~2 is among
 the models already selected in the previous article, since the present oscillation frequencies agree with those published in Paper I within about 5~nHz. 
Model~1 has been now discarded without major doubts, but we decided to report the differences between observed and theoretical frequencies also for this model, to show how a larger set of observations can help to improve the asteroseismic analysis.
This was necessary because, as demonstrated by
 \citet{schunker} and \citet{reese}
  the choice of a mismatched model can lead to erroneous conclusions about the internal rotational profile. 

On the other hand, it was demonstrated in Paper~I and previously found by \citet{deheuvels2014} that
no significant difference is found in the inversion results by using different stellar models, if these models are consistent with each other to first order, which means that they are able to match within the errors both seismic and non-seismic parameters.
In Paper~I it was shown that two selected best-fit models, chosen on the basis of  the $\chi^2$ criterion, produce similar inversion results in the core and in the upper layers.
Those results for two different models not only agreed with each other, but were confirmed by applying different independent methods. 
Some discrepancy
rising from the use of different models has been found only in the regions above the core, as it is shown in Fig.~7 in Paper~I, where the solutions obtained for the two selected models agree within
about 2$\sigma$ errors. \citet{reese} well explained that the reason is that different models might have oscillation modes with different inertia, because of possible different extents of the acoustic and gravity cavities. In the case of red giants, the regions above the core are mostly sounded by modes with mixed g-p character (see the  blue shaded region in Fig. \ref{prop} which will be described below), whose identification is more crucial than modes with dominant p or g behaviour, respectively better trapped in the convective region and in the inner core.
Fortunately, mixed modes of different inertia have very different damping times and profiles in the observed oscillation spectrum.
Thus, the analysis of the observations can provide essential information on
 the gravity or acoustic nature of each detected oscillation mode (see Paper I and Sec. 2) helpful to define
the trapping regions inside the star and, as result, the structure of the model. 
In the following we can proceed by considering with a high degree of confidence only Model~2 as the best-fit model
for the star.
 
\begin{deluxetable}{lccc}
\tabletypesize{\footnotesize}
\tablewidth{0pc}
 \tablecolumns{4}
 \tablecaption{Main parameters for KIC~4448777 and for the best fitting models. 
 \label{tab:fitted}}
 \tablehead{\colhead{} & \colhead{KIC~4448777}& \colhead{Model~1} & \colhead{Model~2}}
\startdata
 $M/{\mathrm M}_{\odot}$& -  &1.02 & 1.13  \\
 Age (Gyr) & -&8.30& 7.24 \\
 $T_{\mathrm{eff}}$ (K) &$4750\pm250\tablenotemark{a}$&  4800 &  4735 \\
 $\log g$ (dex) &$3.5\pm0.5\tablenotemark{a}$&   3.26 &   3.27 \\
 $R/{\mathrm R}_{\odot}$ & - & 3.94 &  4.08 \\
 $L/{\mathrm L}_{\odot}$ & - &7.39 &   7.22 \\
 $Z_{i}$ & - &0.015 &  0.022 \\
$X_{i}$ & - & 0.69 & 0.69 \\
$[\mathrm{Fe/H}]$ & $0.23\pm 0.12 \tablenotemark{a}$ &$-0.04$  & $0.13$\\
$r_{cz}/R$& - & 0.1542 & 0.1448 \\ 
$r_{He}/R$ & - & 0.0075  & 0.0074 \\
$\Delta \nu$ & $16.973\pm0.008$ & $16.893$ & $16.933$
 \enddata
\tablenotetext{a}{Determined by spectroscopic observations \\ (see Paper I)}
 \end{deluxetable}
 
\begin{figure}[!ht]
   \centering
 \includegraphics[height=12cm]{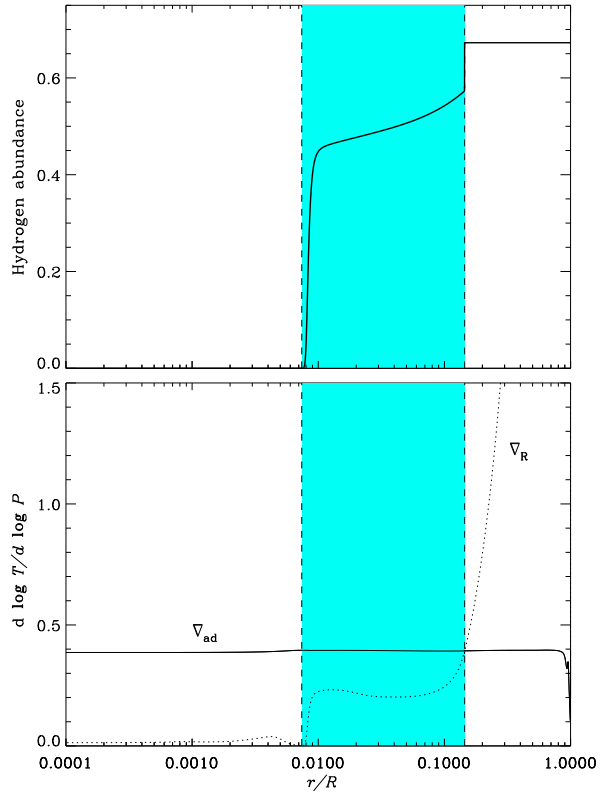}
 \caption{Hydrogen content (upper panel) and temperature gradients (lower panel) in Model~2 of KIC~4448777. The extent of the core and the base of the
 convective envelope located respectively at $r_{He}=0.0074R$  and $r_{cz}= 0.1448R$ are shown by the dashed lines. In the lower panel the solid line shows the adiabatic gradient $\nabla_{ad}$ and the dotted line shows the radiative gradient $\nabla_{R}$. The H-burning shell is shaded in light blue.}
   \label{Xr}
 \end{figure}

\begin{figure}[!ht]
   \centering
 \includegraphics[height=5.5cm]{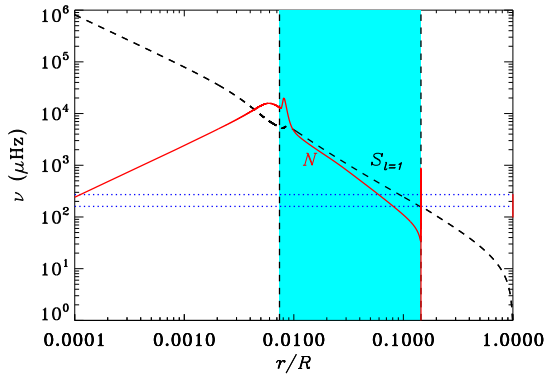}
 \caption{Propagation diagram for Model~2. The red solid-line represents the Buoyancy frequency, while the dashed line indicates the Lamb frequency for $l=1$. The blue dotted lines indicate the range of observed frequencies for KIC~4448777.
 The H-burning shell is shaded in light blue}
   \label{prop}
 \end{figure}

  KIC~4448777, as it was shown in Paper I,  has a degenerate helium core, having  exhausted its central hydrogen and it is still burning hydrogen in the shell. The hydrogen abundance and the temperature gradients as a function of the fractional radius plotted in Fig. \ref{Xr} for Model 2 show the extent of the core with a radius $r_{He}=0.0074R$ and the location of the base of the deep convective zone $r_{cz}=0.1448R$. The convection zone can be easily identified in the  lower panel of Fig. \ref{Xr}   as the region where the radiative gradient is greater than the adiabatic one $\nabla_R>\nabla_{ad}$.
 
 The diagram in Fig. \ref{prop} shows the regions of propagation of the gravity and acoustic modes as
 delimited by the buoyancy and the Lamb frequencies, respectively. The gravity and acoustic cavities appear to interact mostly in the region around the base of the convective zone at $r_{cz}=0.1448R$. This means that this region can be probed by  modes with both gravity-acoustic character: the modes which are commonly observed in red-giant stars. The convective region can be probed mainly by modes better trapped in the acoustic cavity, which behave as pure p modes and have an inertia close to that of radial modes. On the other hand, the He core can be probed with modes mainly trapped in the gravity cavity, with very high inertia.
 
 \section{Results of the asteroseismic inversions}

\begin{figure*}[!ht]
 \begin{center}
\centering
\includegraphics[width=12cm]{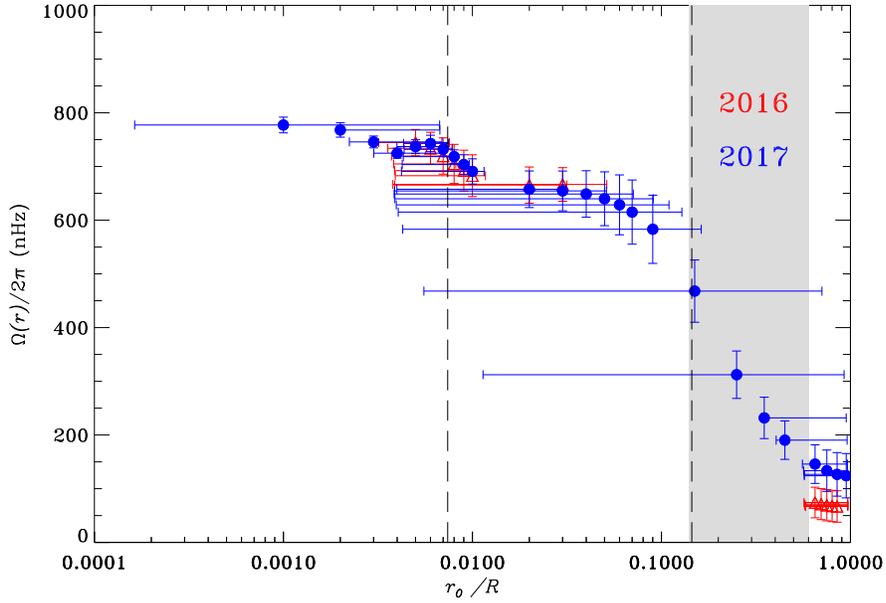}
\end{center}
\caption{Internal rotation of KIC~4448777 at different depths as obtained by the OLA inversion with data of Paper I (flagged by 2016 in red ) and present data (flagged by 2017 in blue).  Vertical error bars are $2\sigma$ of the standard deviations. The dashed lines indicate the  base and the external edge of the H-burning shell. The shaded area marks the region inside the star where the spatial resolution becomes low.}
\label{rot}
\end{figure*}

The inferred rotation rate obtained by applying the Optimally Localized  Averaging (OLA) technique \citep{backus1970} (see Paper I for details) for Model~2 is shown
in Fig. \ref{rot}, where the points indicate the angular velocity against the selected target radii $r_0$. For comparison, the results obtained in Paper I, flagged as 2016, are also indicated. The radial spatial resolution is the interquartile range of the averaging kernels and gives a measure of the localization of the solution.  
Figure \ref{rot} shows that the helium core below $r=0.0074R$ rotates rigidly with an average angular velocity of $\langle{\Omega_c}\rangle=746\pm30$~nHz 
  and  a maximum value of $\Omega_c=777\pm7$~nHz at $r_c=0.001R$.
  
The rotation starts to slow down gradually as the radius increases just entering the inner edge of the hydrogen-burning shell, while  in the interval $0.01 R \leq r\leq 0.05R$, the hydrogen shell seems to rotate nearly at a constant velocity of $\Omega_H=650\pm50$~nHz.
 
As predicted by theory, the convective region decouples from  the core and the angular velocity further decreases (in some way that we cannot efficiently probe) reaching a constant value of  $\Omega_s=124\pm21$~nHz at the surface. This value is slightly higher than that obtained in Paper I, but still compatible with it within the errors. The small difference is due to the  additional dipole modes,  better concentrated in the acoustic cavity and then more suitable to probe these layers, detected in the four-year data set whose splittings have been included in the inversion procedure of the present work.

Similar results can be obtained by inverting the set of rotational splittings by using the Subtractive Optimally Localized Averaging (SOLA) technique \citep{Pijpers1992}, but as already discussed in Paper I, SOLA inversion fails to produce solutions above the He core.

\begin{figure*}[!ht]
\centering
\includegraphics[width=10cm]{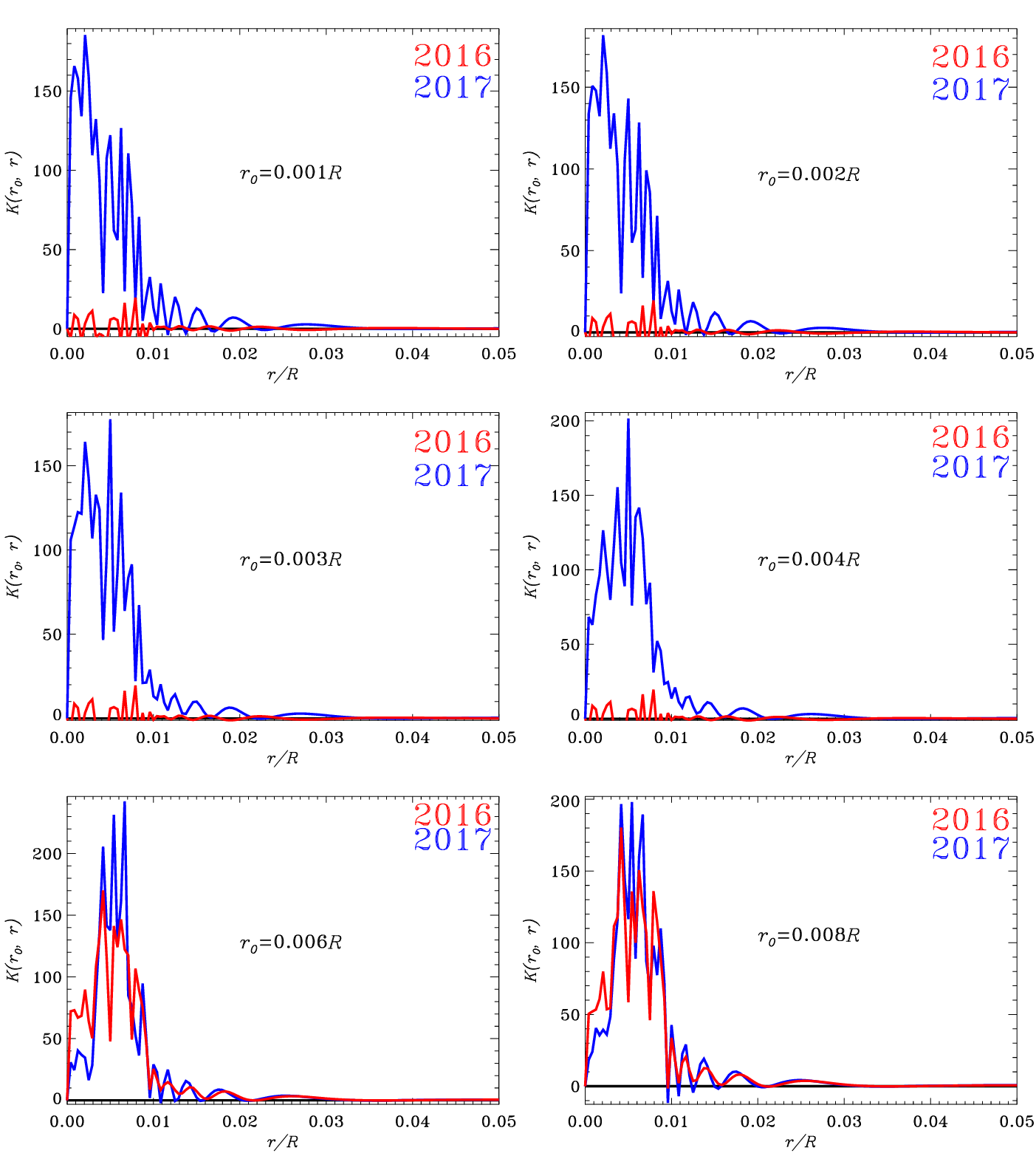}
\caption{Averaging kernels for OLA inversion indicating localization of the solutions inside the He core and part of the H-burning shell as obtained by using present data (blue line) and data of Paper I (red line).}
\label{ker1}
\end{figure*}

\begin{figure*}[!ht]
 \begin{center}
\centering
\includegraphics[width=12cm]{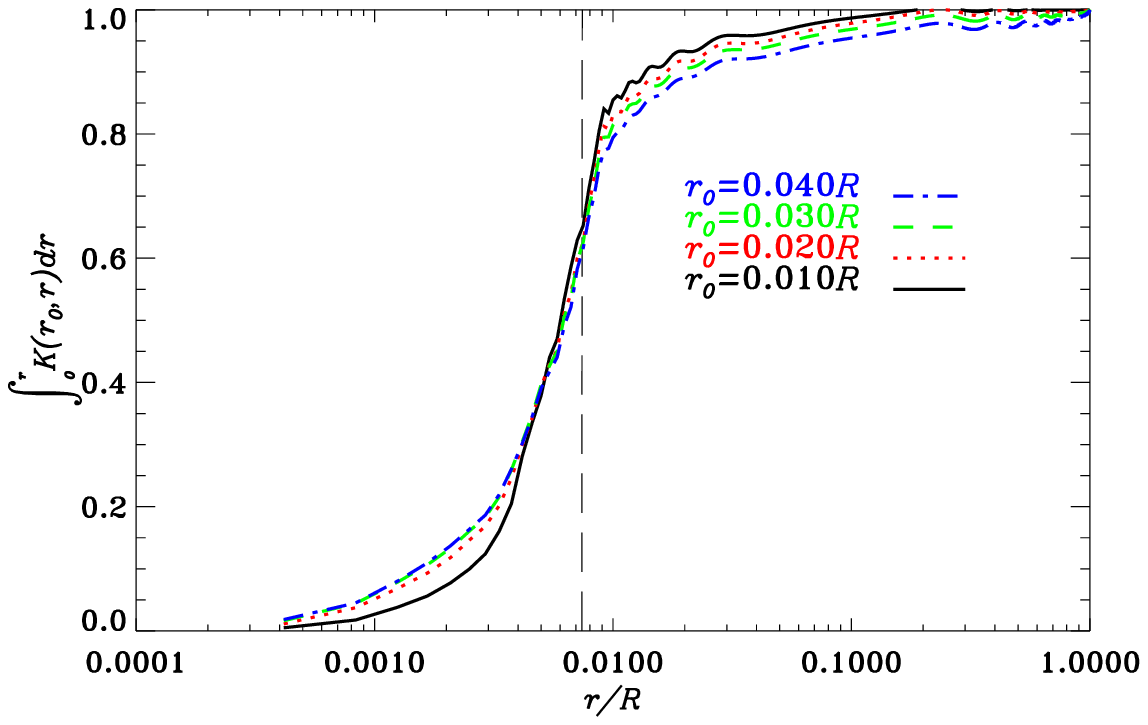}
\end{center}
\caption{OLA cumulative integrals of the averaging kernels centered at different locations in the H-burning shell. The dashed black line indicates the location of the inner edge of the H-burning shell.}
\label{ker2}
\end{figure*}

Figure \ref{ker1}  shows
OLA averaging kernels localized at several target radii $r_0$ inside the core, obtained by adopting a trade-off parameter $\mu=0.001$ which minimizes the propagation of the uncertainties and the spread of the kernels  (see  Sect. 4. of Paper I for details) for the inversion given in Fig. \ref{rot}. 
The new large set of data allows to localize averaging kernels at $r=0.001R$ in the very inner part of the helium core, while previous data set did not allow to find solutions below $r=0.005R$, as shown by the comparison with the
averaging kernels obtained in the analysis of Paper I, also reported in Fig. \ref{ker1}.
In Fig. \ref{ker2} we plot the OLA cumulative integrals of the averaging kernels centered at different locations in the H-burning shell to show the regions of the star where the solutions are most sensitive.  The cumulative kernels are characterized by a strong contamination coming from the He core, but the solutions are still well localized with a percentage of 80\%.

\subsection {The rotational shear between core and envelope}
In order to shed some light over the
 efforts devoted during the last decades to improve the understanding of the physics  and the development of rotational models of low-mass stars, we considered the problem of characterizing the rotational gradient inside KIC~4448777, between the fast rotating core and the slow envelope.
 Intrigued by the work of
 \citet{klion2017}, we compared our results with two possible rotational models which are commonly adopted at present: a rotational model in which angular momentum transport is dominated by local fluid instabilities and the rotational gradient is mainly confined in the hydrogen shell and a model with angular momentum transport dominated by large-scale magnetic fields in which the differential rotation is contained within the convective zone. 
 
 The
application of the OLA inversion technique allowed us to find
evidence for a radial gradient around the base of the convective region. 
As it can be seen in Fig. \ref{rot}, the rotational shear layer appears to lie in good part of the shell where the hydrogen is burning, with a centroid placed close to the base of the convection zone at $r=0.1448$ and
extending from $r=0.05R$ to a not well defined outermost layer above the base of the convective region, which we cannot fully resolve mainly due to the lack of observed modes able to probe into details the envelope, i. e.  modes with low inertia behaving as acoustic modes.  

The method fails to localize the shear with an accuracy better than
$0.08~R$, due to the progressive reduction of the spatial
resolution at increasing distance from the core, which characterizes the observed dipole mixed modes employed in the inversion.
Moreover, both the insufficient spatial resolution and the intrinsic limit of the inversion procedure itself do not allow to infer any information on the gradient of the rotational profile inside the convective region, as found in Paper I and largely discussed
by, e.g., \citet{corbard98}.

\section{Probing the rotation in the convection zone}
 The results obtained raised the natural question about 
the future capability to sound the
regions above the H-burning shell in the hypothesis that additional rotational splittings will be available for the inversion.

In the attempt of answering this question, we used the forward seismological approach as described in Paper I, adopting Model 2 and 
computed artificial splittings from a
 simple fictitious input rotational profile with a rotational gradient occurring inside the H shell:
 \[
 \left\{\begin{array}{ll}
\Omega(r)=750\, \mathrm{nHz} &  r\leq 0.05R;\\
\Omega(r)=120\, \mathrm{nHz}&  r>0.05R.\\
\end{array} \right.
 \] 
Two sets of artificial splittings have been inverted: 1) a set which includes rotational splittings relative to each detected dipolar mode given in Table \ref{tab:frequencies}, which means a total of 20 splittings; 
2) a set which includes rotational splittings for all the theoretical modes with harmonic degree $l=1, 2, 3$, which means a set with 298 splittings. We are aware that the set 2) represents an ideal extreme case.
 
 \begin{figure*}[!ht]
  \begin{center}
\includegraphics[width=12cm]{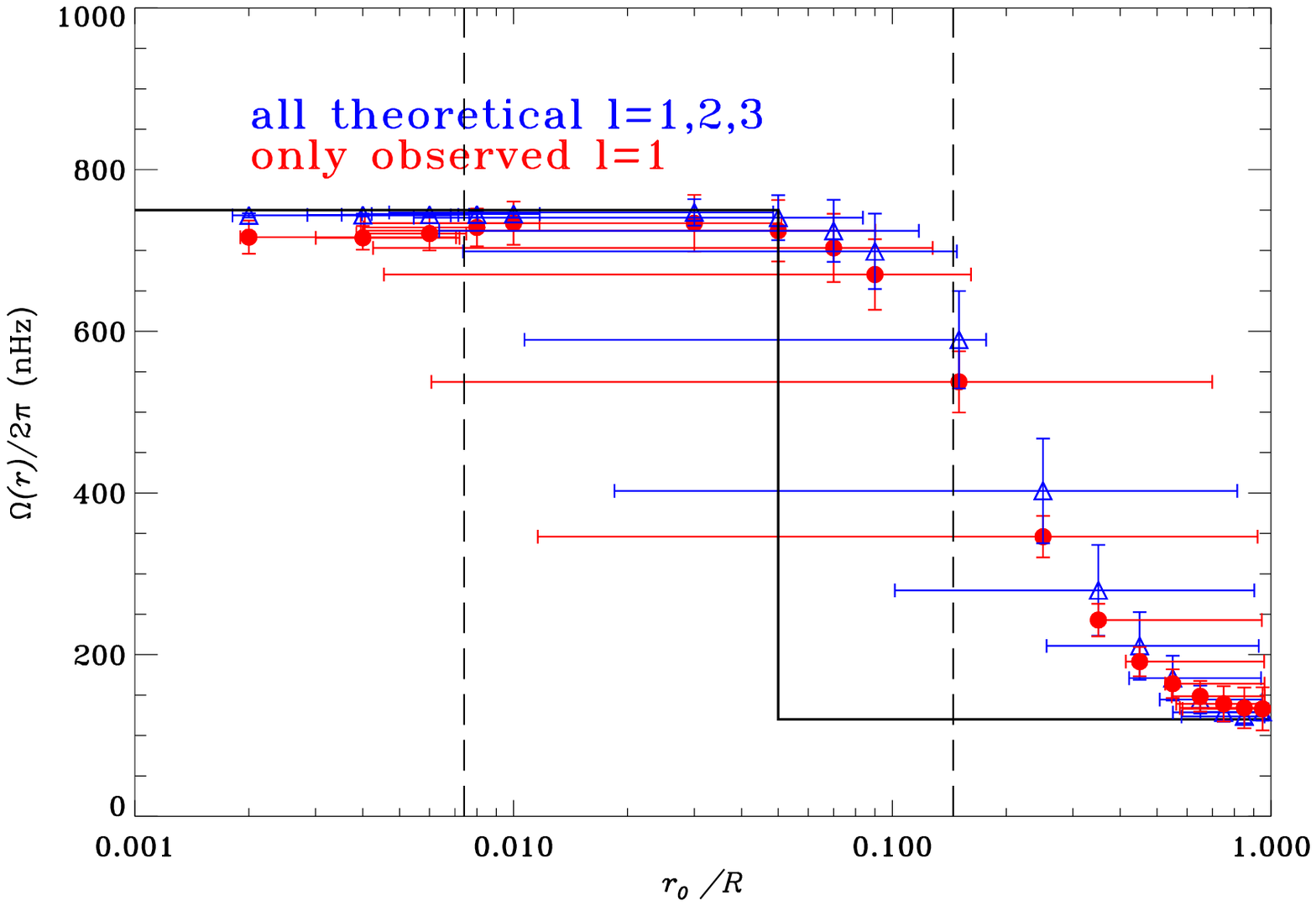}
 \caption{Internal rotation of Model~2 by inversion of two sets of artificial rotational splittings calculated by a forward approach assuming a fictitious internal rotational profile described as a step-like function (solid black line).
   Vertical error bars are $2\sigma$ of the standard deviations. The dashed lines indicate the  base and the external edge of the H-burning shell.} 
 \label{invfin}
  \end{center}
 \end{figure*}
 
Figure \ref{invfin} shows the comparison of the results obtained by inverting set 1) and 2) and we can conclude that:
 by using rotational splittings of low harmonic degree, even with very large set of data, there is no possibility to identify the presence of a steep rotational profile, like that due to the step-like function adopted;  
 the resolution provided by the modes included in the inversion process tends to smooth out the high gradient present in the fictitious rotational profile considered. 
The set which includes 298 splittings of low harmonic degree improves only the radial resolution, but does not help to better distinguish the shape of the rotational profile in the convective zone. 
However, the inversion succeed to place the rotational shear layer and we obtain that the rotation begins to decrease
at $r=0.05R$, where the step is exactly located in our input rotational law.
We would like to point out that we were able also to correctly infer the position of the rotational shear adopting step-like rotational profiles with gradients located at different depths inside the H shell.

It is clear that
the quite low radial resolution in this region depends strongly on the absence - in the inverted set - of modes able to probe the region above $r=0.1 R$. This region, as it is shown in Fig. \ref{prop}, is mainly probed by mixed modes with low inertia, hence with more acoustic character. These modes are too few in the present set of data.

\section{Summary and conclusions}

The results of the present work, obtained by handling a larger set of mixed mode splittings, allowed us to disentangle the details of the rotational internal profile in the red giant KIC~4448777  better than in Paper~I. 

Thanks to the detection of new splittings of mixed modes,  more concentrated inside the He core,  we have been able to reconstruct the angular velocity profile deep into the interior of the star down to $r=0.001R$,  while in  Paper I solutions reached at most  $r=0.005R$. 
The  internal rotational profile of the star appears slightly more complex than predicted by current theories.  In Paper I we found a constant rotational velocity inside the He core and  a smooth decrease from the edge of the He core through the H-burning shell. With the new data we confirm that the He core rotates rigidly from $r=0.001R$ to $r=0.007R$, then the angular velocity slowly decreases as the radius increases first in the layer
of transition between the He core and
at the H-burning shell, around  $r=0.007R$.  

Furthermore,  we found that part of the hydrogen burning  shell appears to rotate rigidly at velocity about 1.15 times lower than the He core, while above $r = 0.05R$ the angular velocity begins to decrease gradually with the increase of the radius. 
Hence, in KIC~4448777 the theoretically predicted shear layer between the fast-spinning core and the slowly rotating envelope is located inside the H shell, above $ r \simeq 0.05 R$, while no definitive conclusion can be drawn on its thickness.  As a matter of interest, in the Sun the 'tachocline' is located just below the convective region at about $ 0.7  \pm 0.005 R_{\odot}$ and has a thickness of about $ 0.05 \pm 0.03 R_{\odot}$  \citep[e.g.,][]{corbard98,dimauro1998,schou1998}.

Our result about the location of the rotational shear layer, agrees with that obtained by applying the method proposed by \citet{klion2017}, which uses the ratio between the minimum rotational splitting of p-dominated modes, min($\delta \nu_p$) and the maximum rotational splitting of g-dominated modes,  max($\delta \nu_g$), measured  in the spectrum of red giants, as an indicator of the location of  the rotational shear layer inside these type of stars.  According to the prescriptions of \citet{klion2017} for less evolved red giants ($R \simeq 4 R_{\odot})$, the value of the ratio which, in our data for KIC~4448777 is 0.348, supports our conclusion that most of the differential rotation is very likely located in the radiative interior of the star, with the rotational profile starting to decrease in the H burning shell. 

In the convective region, decoupled from the radiative interior of the star,  the angular velocity drops down reaching at the surface a value which is about 6 times lower than in the He core.
Unfortunately nothing can be deduced on the profile of the rotational gradient in the convective layer, due to the limited number of modes able to probe efficiently the convective envelope and employed in the inversion procedure. 
Further we theoretically considered, by applying a forward seismological approach and using a
fictitious rotational law,
 the possibility for the future to sound the convective envelope in the red-giant stars by using larger sets of data obtained, i. e., with longer
 time series. We found that a very large set with only low harmonic degree $l\leq3$ will improve only the spatial resolution
 in the convective region,
 while no additional information about the shape of the rotational gradient will be deduced.


We can conclude, that
although the development of evolutionary codes 
has
reached a satisfactory level of accuracy and reliability, there are still  several physical
mechanisms about the internal rotation and the extraction of angular momentum, that are still poorly known.
These physical processes due to their complexity,
sometimes are ignored or just generally
controlled by a number of free parameters.
Hence, we hope that the present asteroseismic results will have important implications for constraining the dominant mechanism of angular momentum transport in this particular phase of evolution, helping to discriminate among the several mechanisms at present considered. This is the case, for example, of the model based on the presence of an internal magnetic field, which has been proposed for modelling the transport of angular momentum inside an evolving star \citep[e.g.,][]{kissin2015}. The present results seem to disfavor this class of models which lead to a flat profile in the radiative zone. Morevover \citet{cantiello} show that models including the Taylor-Spruit dynamo are not able to reproduce the low values of core rotation rates observed in red giants.
For the future, we expect that results on individual stars, like those presented here, will be used to test rotation models, to constrain the variation of the angular velocity within the convective envelope and to test the existence of strong differential rotation across the core-envelope boundary.

\acknowledgements{ We acknowledge the entire Kepler team, whose efforts made these results possible. Funding for the Kepler mission was provided by NASA's Science Mission Directorate.

We are grateful to the anonymous referee
for giving us
the chance to improve this
manuscript.

E.C. is funded by the European Unions Horizon 2020
research and innovation program under the Marie Sklodowska-Curie grant
agreement No. 664931 and by the European Community's Seventh Framework
Programme (FP7/2007-2013) under grant agreement No. 312844
(SPACEINN). B.L.M is grateful for support to National Council for Scientific and Technological Development - CNPq/Brazil}



\end{document}